\documentclass[twocolumn,showpacs,preprintnumbers,amsmath,amssymb]{revtex4}

\usepackage{graphicx}
\usepackage{dcolumn}
\usepackage{bm}

\begin{document}

\preprint{}

\title{
Observation by an Air-Shower Array in Tibet of the Multi-TeV Cosmic-Ray
Anisotropy due to Terrestrial Orbital Motion Around the Sun
}

\author{
{M.~Amenomori},$^1$
{S.~Ayabe},$^2$
{S.W.~Cui},$^3$
{Danzengluobu},$^4$
{L.K.~Ding},$^3$
{X.H.~Ding},$^4$
{C.F.~Feng},$^5$
{Z.Y.~Feng},$^6$
{X.Y.~Gao},$^7$
{Q.X.~Geng},$^7$
{H.W.~Guo},$^4$
{H.H.~He},$^3$
{M.~He},$^5$
{K.~Hibino},$^8$
{N.~Hotta},$^9$
{Haibing~Hu},$^4$
{H.B.~Hu},$^3$
{J.~Huang},$^{10}$
{Q.~Huang},$^6$
{H.Y.~Jia},$^6$
{F.~Kajino},$^{11}$
{K.~Kasahara},$^{12}$ 
{Y.~Katayose},$^{13}$
{C.~Kato},$^{14}$
{K.~Kawata},$^{10}$
{Labaciren},$^4$
{G.M.~Le},$^{15}$
{J.Y.~Li},$^5$
{H.~Lu},$^3$
{S.L.~Lu},$^3$
{X.R.~Meng},$^4$
{K.~Mizutani},$^2$
{S.~Mori},$^{14}$
{J.~Mu},$^7$
{K.~Munakata},$^{14}$
{H.~Nanjo},$^1$
{M.~Nishizawa},$^{16}$
{M.~Ohnishi},$^{10}$
{I.~Ohta},$^9$
{H.~Onuma},$^2$
{T.~Ouchi},$^{10}$
{S.~Ozawa},$^9$
{J.R.~Ren},$^3$
{T.~Saito},$^{17}$
{M.~Sakata},$^{11}$ 
{T.~Sasaki},$^8$
{M.~Shibata},$^{13}$
{A.~Shiomi},$^{10}$
{T.~Shirai},$^8$
{H.~Sugimoto},$^{18}$
{M.~Takita},$^{10}$
{Y.H.~Tan},$^3$
{N.~Tateyama},$^8$
{S.~Torii},$^8$
{H.~Tsuchiya},$^{10}$
{S.~Udo},$^2$
{T.~Utsugi},$^8$
{B.S.~Wang},$^3$
{H.~Wang},$^3$
{X.~Wang},$^2$
{Y.G.~Wang},$^5$
{H.R.~Wu},$^3$
{L.~Xue},$^5$
{Y.~Yamamoto},$^{11}$
{C.T.~Yan},$^{3}$
{X.C.~Yang},$^7$
{S.~Yasue},$^{14}$
{Z.H.~Ye},$^{15}$
{G.C.~Yu},$^6$
{A.F.~Yuan},$^4$
{T.~Yuda},$^{10}$
{H.M.~Zhang},$^3$
{J.L.~Zhang},$^3$
{N.J.~Zhang},$^5$
{X.Y.~Zhang},$^5$
{Y.~Zhang},$^3$
{Zhaxisangzhu}$^4$ and
{X.X.~Zhou}$^6$ \\
( Tibet AS$\gamma$ Collaboration )
\\
{\it
$^1$  Department of Physics, Hirosaki University, Hirosaki 036-8561, Japan \\
$^2$  Department of Physics, Saitama University, Saitama 338-8570, Japan \\
$^3$  Institute of High Energy Physics, Chinese Academy of Sciences, Beijing 100039, China \\
$^4$  Department of Mathematics and Physics, Tibet University, Lhasa 850000, China \\
$^5$  Department of Physics, Shandong University, Jinan 250100, China \\
$^6$  Institute of Modern Physics, South West Jiaotong University, Chengdu 610031, China \\
$^7$  Department of Physics, Yunnan University, Kunming 650091, China \\
$^8$  Faculty of Engineering, Kanagawa University, Yokohama 221-8686, Japan \\
$^9$  Faculty of Education, Utsunomiya University, Utsunomiya 321-8505, Japan \\
$^{10}$ Institute for Cosmic Ray Research, the University of Tokyo, Kashiwa 277-8582, Japan \\
$^{11}$ Department of Physics, Konan University, Kobe 658-8501, Japan \\
$^{12}$ Faculty of Systems Engineering, Shibaura Institute of Technology, Saitama 330-8570, Japan \\
$^{13}$ Faculty of Engineering, Yokohama National University, Yokohama 240-8501, Japan \\
$^{14}$ Department of Physics, Shinshu University, Matsumoto 390-8621, Japan \\
$^{15}$ Center of Space Science and Application Research, Chinese Academy of Sciences, Beijing 100080, China \\
$^{16}$ National Institute for Informatics, Tokyo 101-8430, Japan \\
$^{17}$ Tokyo Metropolitan College of Aeronautical Engineering, Tokyo 116-0003, Japan \\
$^{18}$ Shonan Institute of Technology, Fujisawa 251-8511, Japan
}
}

\date{\today}

\begin{abstract}
We report on the solar diurnal variation of the galactic cosmic-ray intensity observed by the Tibet~III air shower array during the period from 1999 to 2003.
In the higher-energy event samples (12~TeV and 6.2~TeV), the variations are fairly consistent with the Compton-Getting anisotropy due to the terrestrial orbital motion around the sun,
while the variation in the lower-energy event sample (4.0~TeV) is inconsistent with this anisotropy.
This suggests an additional anisotropy superposed at the multi-TeV energies, e.g. the solar modulation effect.
This is the highest-precision measurement of the Compton-Getting anisotropy ever made.
\end{abstract}

\maketitle

\section{\label{sec:level1}Introduction}
The galactic anisotropy of the cosmic-ray intensity is expected to carry information about the origin and the propagation mechanism of the galactic cosmic rays, as it reflects the magnetic field in space through which cosmic rays have traveled.
The anisotropy can be observed as the daily variation of cosmic-ray intensity recorded by the ground based detector in the sidereal time (sidereal daily variation).
The amplitude of the sidereal daily variation so far reported is as small as 0.1$\%$ or less, while the amplitude of the temporal variation of the air-shower event trigger rate amounts to a few$\%$ at the experimental site.
Therefore, we have to eliminate the temporal variation which is due mostly to the atmospheric pressure and temperature effects.

We still have no reliable theoretical constraints for the sidereal daily variation due to the galactic anisotropy.
This also makes it difficult to evaluate the systematic error contained in the observed variation.
Accordingly, we first analyze the daily variation in the solar time, which is expected from the so-called the Compton-Getting (C-G) anisotropy  \cite{cg-1935} due to the earth's orbital motion around the sun.
As this effect is predicted based on a reliable theory, the positive observation of this effect will assure the reliabilities of both the measurement and analysis.
When a cosmic-ray detector on the earth moves with respect to the rest frame of the cosmic-ray plasma, the fractional intensity enhancement due to the C-G anisotropy is expressed, as Eq.~(\ref{eq})
\begin{equation}
\frac{\Delta I}{<I>}~=~(\gamma+2)~\frac{v}{c}~\cos\theta \; ,
\label{eq}
\end{equation}
with $I$ denoting the cosmic-ray intensity,
$\gamma$ the power-law index of the cosmic-ray energy spectrum,
$v/c$ the ratio of the detector's velocity to the speed of light,
and $\theta$ the angle between the arrival direction of cosmic rays and the direction of motion of detector \cite{cg-1968}.
The vertical viewing detector on the earth moving along the circular orbit around the sun scans various directions in space as the earth spins and records the C-G anisotropy as the solar diurnal variation in its counting rate with a maximum at 6:00 hours in the local solar time.
The amplitude of this variation is calculated to be as small as 0.05$\%$ or less, 
depending on the geographic latitude of the experimental site. 

The solar diurnal variation also can be caused by the anisotropy due to the solar modulation of galactic cosmic rays in the heliosphere \cite{Axford-1965}.
Yasue et al. \cite{matsushiro-1991} reported from their underground muon experiments that the anisotropy due to the solar modulation superposed on the C-G anisotropy extended up to several 100 GeV (see also \cite{munakata-icrc25}).

Above 10~TeV energies, there have been only a few positive observational results reported on the solar diurnal variation due to the C-G anisotropy
\cite{baksan-eas1981} \cite{nagashima-1989} \cite{eas-top-1996},
because of very small amplitude of the variation.
Cutler and Groom reported in 1986 the first clear signature of the C-G anisotropy for multi-TeV cosmic rays ($\sim$ 1.5~TeV) in the solar diurnal variation \cite{utah-1986}.
The variation reported by them was in reasonable agreement with the sinusoidal curve expected from the C-G anisotropy, while the maximum phase of the curve deviated from 6:00 hours by $+2$ hours at 2~$\sigma$ significance.
They attributed the deviation to the meteorological effect on the underground muon intensity.

Therefore, the solar diurnal variation is so far presumed to be free from the solar modulation effect at TeV energies.
We will report on the solar diurnal variation due to the C-G anisotropy observed by the Tibet~III air shower array.
The analysis of the sidereal anisotropy will be published elsewhere.

\section{Experiment}

The Tibet air shower experiment has been successfully operated at Yangbajing (90.522$^\circ$~E, 30.102$^\circ$~N, 4300~m above sea level) in Tibet, China since 1990.
The array constructed first in 1990 was gradually upgraded by increasing the number of counters \cite{tibet} \cite{tibet-crab} \cite{tibet2000} \cite{tibet2002},
and then the Tibet~III array, used in the present analysis, was completed in the late fall of 1999.
This array consists of 533 scintillation counters of 0.5~m$^2$ each placed on a 7.5~m square grid with an enclosed area of 22,050~m$^2$ and each viewed by a fast-timing (FT) photo-multiplier tube.
A 0.5~cm thick lead plate is put on the top of each counter in order to increase the array sensitivity by converting $\gamma$ rays into electron-positron pairs.

An event trigger signal is issued when any fourfold coincidence occurs in the FT counters recording more than 0.6 particles, resulting in the trigger rate of about 680 Hz at a few-TeV threshold energy.
We collected $5.4 \times 10^{10}$ events by the Tibet~III array during 918 live days from November, 1999 to November, 2003.
After some simple data selections (software trigger condition of any fourfold coincidence in the FT counters recording more than 0.8~particles in charge, zenith angle of arrival direction $< 45^\circ$ , air shower core position located in the array, etc), $3.0 \times 10^{10}$ events remain for further analysis.

The pointing accuracy (0.02$^\circ$) and angular resolution (0.9$^\circ$) of the Tibet~III array can be directly checked by monitoring the Moon's shadow in the cosmic-ray flux at multi-TeV energies \cite{tibet2000} \cite{kawata2003}.

The performance of the Tibet~III array is also examined by means of a full Monte Carlo (MC) simulation in the energy range from 0.3 to 1000~TeV.
We used the CORSIKA version 6.004 code \cite{corsika} and QGSJET model \cite{qgsjet} for the generation of air shower events and the EPICS UV7.24 code \cite{epics} for the detector simulation of shower particles with scintillation counters, respectively.
Primary cosmic-ray particles are sampled from the energy spectrum made by a compilation of direct observational data. 
The primary cosmic-ray energy is estimated by $\Sigma \rho_{\rm FT}$ which is the sum of the number of particles/m$^{2}$ for each FT counter.
According to the result of the simulation, $\Sigma \rho_{\rm FT}~=~100$ corresponds approximately to 10~TeV primary cosmic-ray energy \cite{tibet2001a}.

\section{Analysis}

The selected air shower events are subsequently histogrammed in hourly bins in the solar local time (365~cycle/year), according to event time, incident direction and air shower size of each event.
To check the seasonal change in the daily variation, we obtained the histogram for each month and corrected it for the observation live time varying month to month.

The daily and yearly event rates vary by $\pm 2\%$ and $\pm 5\%$ \cite{kawata2003}, respectively, due mostly to the meteorological effect.
To eliminate these temporal variations and discuss the daily variation with very small amplitude (0.05$\%$ or less), we adopt the following East$-$West subtraction method.
We first obtain the daily variation in the solar time for each of `East' and `West' (E- and W-) incident events referring to the geographical longitude of the incident direction of each event and then subtracts the variation in the W-incident events from that in the E-incident events.
Dividing this difference by the hour angle separation between the mean E- and W-incident directions averaged over the E- and W-incident events, we finally reach the ``differential'' variation at solar time frame.
This method \cite{nagashima-1989} largely cancels out the meteorological effect and the possible detector biases, which are expected to produce common variations for both the E- and W-incident events.

A possible drawback of this method is that we obtain only the ``differential'' form of the physical variation and we have to reconstruct the physical variation by ``integrating'' the ``differential'' variation with respect to the local time.
This also makes the direct error estimation difficult.
Hereafter, we will make the statistical argument on the basis of the ``differential'' variation ``$D(t)$'' and compare the physical variation ``$R(t)$'' with model curves obtained by integrating the best-fitted curve to ``$D(t)$''.

The data are then divided into the 3 data samples according to the representative primary energy of 4.0, 6.2 and 12~TeV.
Each representative energy is calculated as the mode value of the logarithmic energy of each event by the MC simulation.

The expected event rate at our experimental site is calculated by considering the effective area of Tibet~III and the C-G anisotropy given in Eq.~\ref{eq} with $\gamma$ of 2.7.
The expected ``$D(t)$'' is then calculated by applying the E$-$W method and compared with the data.

\section{Results and Discussions}

\begin{figure}[ht]
\begin{center}
\includegraphics[width=1.0\linewidth]{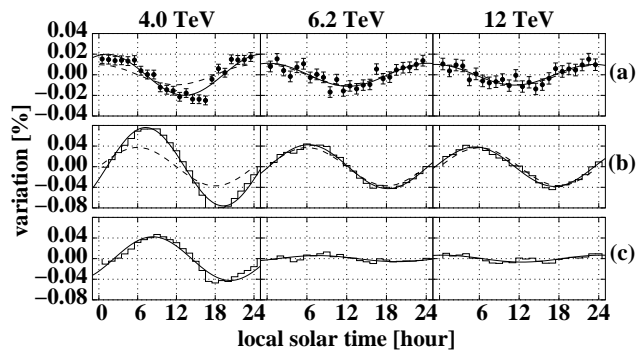}
\caption{
Average solar daily variations recorded in the 3 data samples with representative energies of 4.0, 6.2 and 12~TeV.
The expected variations due to the C-G anisotropy are shown by broken lines, while the sinusoidal curves best-fitted to the data are shown by solid lines.
From the top shown are the ``differential'' variations ``$D(t)$''s (a), the physical variations ``$R(t)$''s (b) and the differences between ``$R(t)$'' and the expected variation (c), respectively.
The error bars are statistical.
}
\label{sol-fit}
\end{center}
\end{figure}

Figure~\ref{sol-fit} shows the average solar daily variations observed by the Tibet~III together with the sinusoidal curves best-fitted to the data.
Note that the variations in Fig.~\ref{sol-fit}(a) are the ``differential'' variations (``$D(t)$''s) and the maximum phase in each panel is shifted earlier by 6~hours (1/4 cycles) from the  corresponding actual daily variation.
The amplitudes in this panel are also $\pi/12$ times as small as those of the actual variations.

The $\chi^2$-fitting results for the 3 data samples, assuming a sinusoidal curve, are summarized in Table~\ref{table}.
The variations of the higher-energy data samples (12~TeV and 6.2~TeV) are consistent with those expected from the C-G anisotropy, while the variation of the lower-energy data sample (4.0~TeV) statistically deviates from the expected C-G curve at 5.3~$\sigma$ significance in phase and at 8.3~$\sigma$ significance in amplitude, respectively.
Compared with the expected C-G anisotropy, the $\chi^2$ value for the 4.0~TeV data sample is calculated to be 90.5/24 degree of freedom (d.o.f.), suggesting that the variation of this data sample is statistically inconsistent with the C-G anisotropy.
It is noted here that the variation in the 6.2~TeV data sample is consistent with the smooth interpolation between those in the 4.0~TeV and 12~TeV data samples as well as with the C-G anisotropy within statistics.

\begin{table}[h]
\begin{center}
\begin{tabular}{c|c|c|c|c|c}
Energy & \multicolumn{2}{c|}{${\cal A} [\times10^{-3}\%]$} & \multicolumn{2}{c|}{$\phi$ [hour]} & \\ [0ex]
\cline{2-6} \\ [-2.5ex]
[TeV]	& C-G & data & C-G & data & $\chi^2$/d.o.f.\\ [0ex]
\hline
4.0	&     & 19.9$\pm$1.2 &     & 7.25$\pm$0.24 & 32.0/22\\ [0ex]
6.2	& 9.7 & 11.1$\pm$1.5 & 6.0 & 6.23$\pm$0.51 & 19.6/22\\ [0ex]
12	&     & 10.0$\pm$1.6 &     & 5.34$\pm$0.62 & 11.9/22\\ [0ex]
\hline
\end{tabular}
\caption{Amplitudes ${\cal A}$ and phases $\phi$ of the diurnal variations of the $\chi^2$ -fitted results assuming a sinusoidal curve, where the error bars are statistical in the data.
Errors in the expected C-G anisotropy are negligible.
}
\label{table}
\end{center}
\end{table}

\begin{figure}[ht]
\begin{center}
\includegraphics[width=1.0\linewidth]{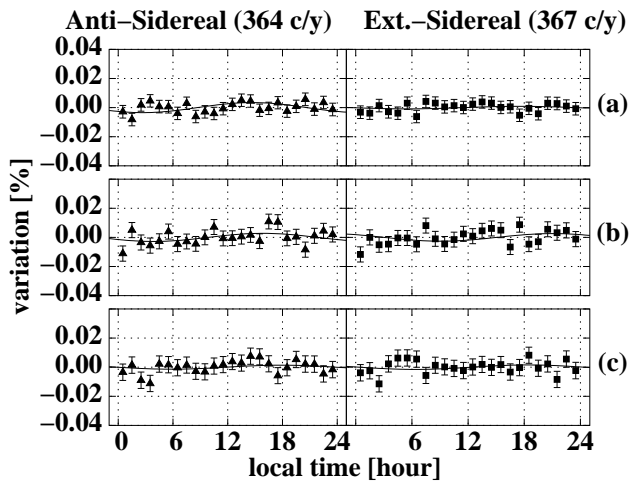}
\caption{
The ``differential'' variations ``$D(t)$''s in the local anti-sidereal time (left) and extended-sidereal time (right).
From the top shown are the average variations recorded in 4.0~TeV (a), 6.2~TeV (b) and 12~TeV (c) data samples, respectively.
The amplitudes of sinusoidal curves best-fitted to all data (3.0 $\times$10$^{10}$ events) are 0.0021$\pm$0.0008$\%$ in the anti-sidereal time and 0.0015$\pm$0.0008$\%$ in the extended-sidereal time, respectively.
The error bars are statistical.
}
\label{asid-cmp}
\end{center}
\end{figure}

A significant spurious variation in the solar time can be produced from the seasonal change of the sidereal daily variation due to the galactic anisotropy, as the average variation in sidereal time is expected to be a few times larger than the solar daily variation due to the C-G anisotropy \cite{munakata-2001}.
Figure~\ref{asid-cmp} shows ``$D(t)$''s distributions in the anti-sidereal (364~c/y) and extended-sidereal (367~c/y) time frames, which are both statistically insignificant.
The insignificant variation (0.0021$\pm$0.0008$\%$) in anti-sidereal time ensures that the seasonal change of the solar daily variation (365~c/y) is negligible.
The insignificant variation (0.0015$\pm$0.0008$\%$) in the extended-sidereal time, on the other hand, supports that the seasonal change of the sidereal daily variation (366~c/y) is also negligible.
These results indicate that the spurious variation contained in the average variation is small.
We thus estimate that the contamination due to the spurious variation might be less than 20$\%$ of the C-G anisotropy.

The 4.0~TeV data sample in Fig.~\ref{sol-fit} is further divided into the 3 sub-samples according to the representative energies to check possible trigger threshold biases.
We confirmed that each of these 3 sub-samples is statistically consistent with one another and the trigger threshold biases are unlikely.

To compare with the result by Cutler and Groom \cite{utah-1986}, which is integrated above the muon threshold energy at 128~GeV and is now the  only one available at multi-TeV energies, we integrate the 4.0~TeV, 6.2~TeV and 12~TeV data samples and give a fit to it by assuming a sinusoidal curve again.
The amplitude normalized by the C-G anisotropy in our work becomes 1.28$\pm$0.08 (stat.) which is not inconsistent with 0.73$\pm$0.21 (stat.) obtained by Cutler and Groom \cite{utah-1986} within statistics.
Thus, the deviation from the C-G anisotropy measured by our observation in the 4.0~TeV data sample may be diluted by the energy integration.
With our high statistics, we can afford to demonstrate the energy dependence of the solar diurnal anisotropy at multi-TeV energies for the first time.

In conclusion, we clearly observe the C-G anisotropy in the 12~TeV and 6.2~TeV data samples.
This is the most precise measurement of the anisotropy ever made.
The measurement also suggests an additional anisotropy superposed on the C-G anisotropy in the 4.0~TeV data sample, for instance, the anisotropy due to the solar modulation effect.
If this extra anisotropy is actually due to the solar modulation, the observed deviation from the C-G anisotropy may vary according to the solar activity changing in every 11 years.
As the data used in the present analysis cover only the solar maximum period, we should examine this hypothesis by continuing the observation for a full solar activity cycle.
The data during the next solar minimum around 2006 might be especially interesting.
Furthermore, it may be also very useful to lower the energy threshold of the array down to sub-TeV region for a better understanding of the phenomenon.
\\

{
This work is supported in part by Grants-in-Aid for Scientific Research on Priority Areas (MEXT) and Scientific Research (JSPS) in Japan, and the Committee of the Natural Science Foundation and the Chinese Academy of Sciences in China.
}

\bibliography{prl-cg}

\end{document}